\documentclass[floatfix,prl,twocolumn,letterpaper,nobalancelastpage]{revtex4-2}

\usepackage{mycommands}

\renewcommand{\section}[1]{\textit{#1}. ---}

\begin{document}
    
    \title{Electric Field Tunable Band Gap in Commensurate Twisted Bilayer Graphene}
    
    \author{Spenser Talkington}
    \affiliation{Department of Physics and Astronomy, University of Pennsylvania, Philadelphia, Pennsylvania 19104, USA}
    \email{spenser@upenn.edu}
    
    \author{Eugene J. Mele}
    \affiliation{Department of Physics and Astronomy, University of Pennsylvania, Philadelphia, Pennsylvania 19104, USA}
    
    \date{\today}
    \begin{abstract}
        Bernal bilayer graphene exhibits a band gap that is tunable through the \textit{infrared} with an electric field. We show that sublattice odd commensurate twisted bilayer graphene (C-TBG) exhibits a band gap that is tunable through the \textit{terahertz} with an electric field. We show that from the perspective of terahertz optics the sublattice odd and even forms of C-TBG are ``inflated" versions of Bernal and AA stacked bilayer graphene respectively with energy scales reduced by a factor of 110 for the $21.79^\circ$ commensurate unit cell. This lower energy scale is accompanied by a correspondingly smaller gate voltage, which means that the strong-field regime is more easily accessible than in the Bernal case. Finally, we show that the interlayer coherence energy is a directly accessible experimental quantity through the position of a power-law divergence in the optical conductivity.
    \end{abstract}
    
    \maketitle
        
       Superlattice heterostructures provide a versatile material  for realizing exotic low energy electronic physics. A prototypical example is \textit{small}-twist magic angle twisted bilayer graphene  which has attracted the most attention with an observed superconducting phase \cite{cao2018unconventional,yankowitz2019tuning} and many other correlated electron phenomena \cite{andrei2020graphene}. More generally twisted bilayer graphene is a versatile platform that can exhibit exotic physics at \textit{large}-twist angles as well. This has been recognized very recently in the observation of the quantum anomalous Hall effect \cite{geisenhof2021quantum} and superconductivity
        \cite{zhou2022isospin} in Bernal twisted bilayer graphene.
       
        It has been known for some time that the application of a perpendicular electric field to Bernal bilayer graphene produces a field tunable band gap through the \textit{infrared} spectrum \cite{mccann2006asymmetry,min2007ab,castro2007biased,mak2009observation}. Meanwhile, large-twist angle bilayer graphene exhibits commensurate structures with band structures that have the same symmetries as primitive Bernal and AA stacked bilayer graphene, but with a reduced interlayer coherence energy scale \cite{shallcross2008quantum}. This motivates our study and its prediction that sublattice-exchange (SE) odd  commensurate twisted bilayer graphene will exhibit an electric field tunable band gap through the \textit{terahertz} spectrum.
        
        While the electronic structure of large-twist angle bilayer graphene is often treated as that of two independent Dirac cones \cite{dos2007graphene,neto2009electronic,rozhkov2016electronic}, this approximation is only true at large energy scales. At small energy scales in commensurate twisted bilayer graphene, there is a residual interlayer coherence resulting from Bragg scattering between Dirac cones  of the two layers by a reciprocal lattice vector \cite{mele2010commensuration,mele2012interlayer,weckbecker2016low}. This interlayer coherence leads to electronic structures in SE-odd bilayer graphene that emulate the quadratic band touching of Bernal bilayer graphene, and that lead to a band gap in SE-even bilayer graphene that is supports topological crystalline insulator edge states \cite{kindermann2015topological}. We show that this interlayer coherence is directly accessible experimentally through the frequency of a power-law divergence in the optical conductivity.
        
        While the linear optical \cite{tabert2013optical,moon2013optical,stauber2013optical}, circular dichroic \cite{kim2016chiral,morell2017twisting,stauber2018chiral,addison2019twist}, and nonlinear optics \cite{zuber2021nonlinear,zuber2021twist}, of bilayer graphene have been studied extensively in the visible and infrared spectra, terahertz studies of large-angle samples with uniform twist angles are missing. A decade ago, Zou \textit{et al}. studied the terahertz optical response of bilayer graphene in Ref. \onlinecite{zou2013terahertz}, but these measurements averaged over samples with many twist angles. The recent interest in moir\'e heterostructures has been led to an increase in sample quality. This increased sample quality could be used to experimentally access flat bands at large twist angles \cite{kindermann2012zero,pal2019emergent,scheer2022magic}, topological crystalline insulator edge states \cite{kindermann2015topological}, or as we suggest here to create a semiconductor with a gap tunable through the terahertz region with an electric field.
    
    \section{Commensuration in Twisted Bilayer Graphene}\label{section:models}
        \begin{figure*}
            \centering
            \includegraphics[width=\linewidth]{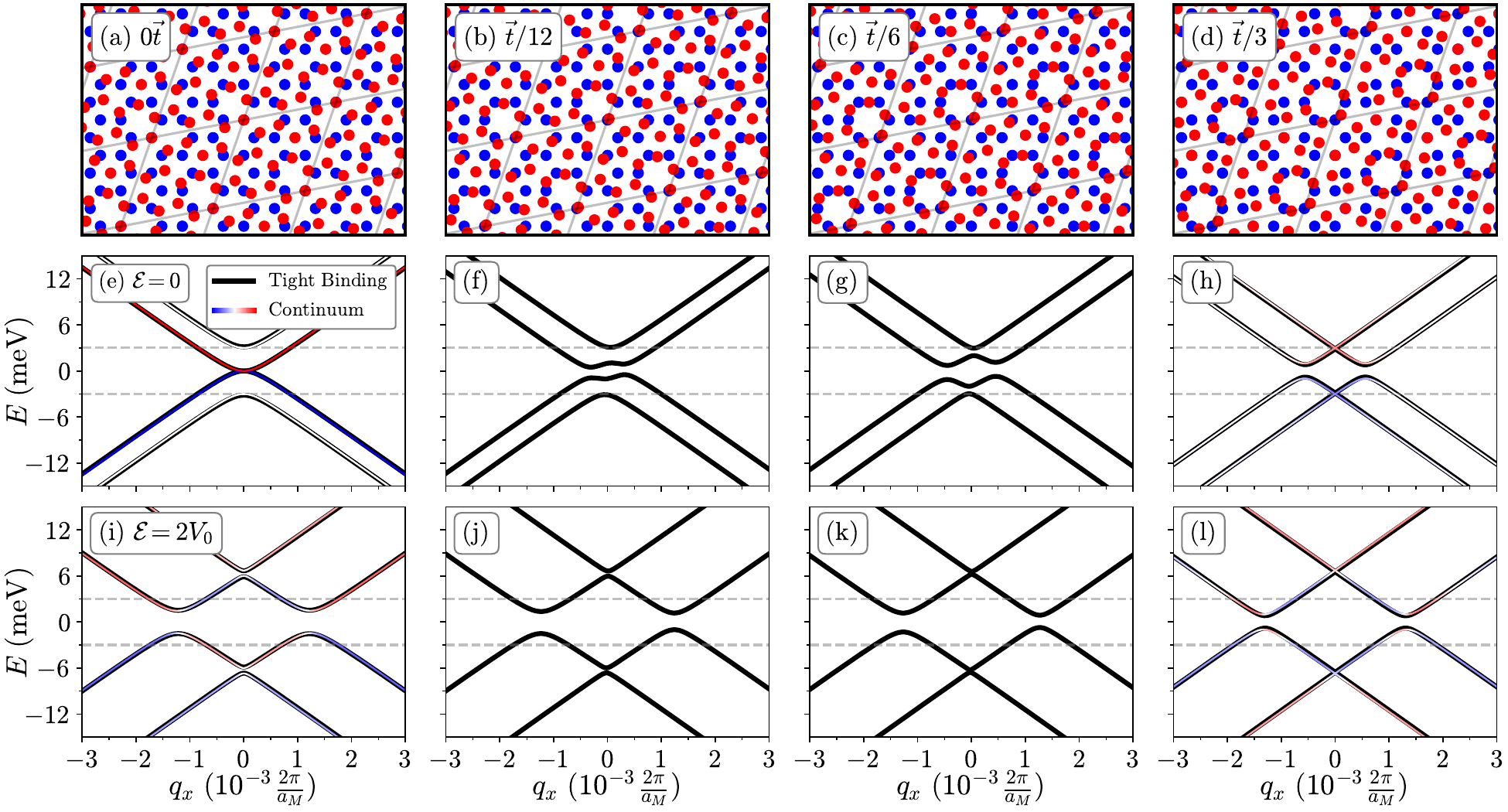}
            \caption{\label{fig:structures}
            Real-space crystal structure and momentum-space band structure of the 21.79$^\circ$ twisted bilayer graphene (TBG) for interlayer various shift vectors. The band structure near the Fermi energy is qualitatively the same AB and AA stacked graphene bilayers but with an energy scale that is two orders of magnitude smaller. The structures vary from a gapless structure (SE-odd) with a quadratic band touching at zero shift to a gapped structure with linear band crossings above and below the Fermi energy (SE-even) at $\vec{t}/3$ shift, where $\vec{t}=(a_{M,1}+a_{M,2})/7$. Here the dashed gray line indicates the interlayer coherence scale, $V_0=3$ meV, a scale which is two orders of magnitude smaller than the scale in AB and AA bilayers. This small energy scale makes the strong-field limit much easier to obtain: to realize layer potential energies of $\pm \mathcal{E}=\pm 2V_0$ requires an electric field strength of just 0.0358 V/nm. In the strong-field limit, these materials exhibit a band inversion (colored by $c=\frac{1}{2}(1+|\langle 3_\text{odd}(0)|\psi(\mathcal{E})\rangle|^2-|\langle 2_\text{odd}(0)|\psi(\mathcal{E})\rangle|^2)$). This inversion can be understood as the layer Dirac cones being separated in the strong-field limit, and then inverting when the interlayer coherence dominates the electric potential.
            (columns): structures as a function of shift, (row 1): crystal structures, (row 2): band structures, (row 3): band structures in an electric field.}
        \end{figure*}
        In twisted bilayer graphene, commensuration occurs and finite unit cells form at twist angles of $\theta(m,n)=\text{Arg}[(me^{-i\pi/6}+ne^{i\pi/6})/(ne^{-i\pi/6}+me^{i\pi/6})]$ for integers $m$, $n$ \cite{campanera2007density}. Of these, $(m,n)=(1,2)$ and $(1,4)$, corresponding to $\theta=30\mp 8.213^\circ$, have the smallest unit cells and largest interlayer coherence \cite{shallcross2008quantum}. Twist angles $\theta$ and $60^\circ-\theta$ correspond to structures with equal unit cell areas, but opposite sublattice-exchange (SE) parities \cite{mele2010commensuration}. SE-odd structures have points with $C_3$ symmetry, while SE-even structures have points with $C_6$ symmetry. As shown in Fig. \ref{fig:structures} panels (a-d), SE-odd and SE-even structures are also related by interlayer lattice translations.
        
        Commensuration ensures that the layer Dirac cones are separated by a reciprocal lattice vector \cite{mele2010commensuration}, which leads to an interlayer coherence that changes the low energy behavior from that of two uncoupled Dirac cones with linear band crossings to quadratic band touchings in the SE-odd case and a gapped structure in the SE-even case \cite{mele2010commensuration}. These SE-odd structure at $21.79^\circ$ is identical to the Bernal graphene bilayer but with an energy scale that is two orders of magnitude smaller.
        DFT calculations find that this interlayer coherence scale, $V_0$ is $\sim 4$ meV \cite{shallcross2008quantum}, although this is at the energy resolution limit of DFT. Using a different $V_0$ linearly scales the results with no qualitative changes. Using a two-center Slater-Koster type tight-binding model for hopping between the $p_z$ orbitals of the carbon atoms \cite{trambly2010localization,moon2013optical}, we find the low energy band structure of these systems in the absence of an applied electric field $\mathcal{E}=0$ and show the results in Fig. \ref{fig:structures} panels (e-h). The low energy states are near the $K$ and $K'$ points of the superlattice Brillouin zone and we expand in $\bm{q}=\bm{k}-\bm{k}_K$ about $K$. Using this tight-binding model we find an interlayer coherence scale of $V_0=3.0$ meV, which we use for the remainder of the paper \footnote{See Supplemental Material at [URL will be inserted by publisher] for parameters for the tight-binding model.}.

\begin{figure*}
    \centering
    \includegraphics[width=\linewidth]{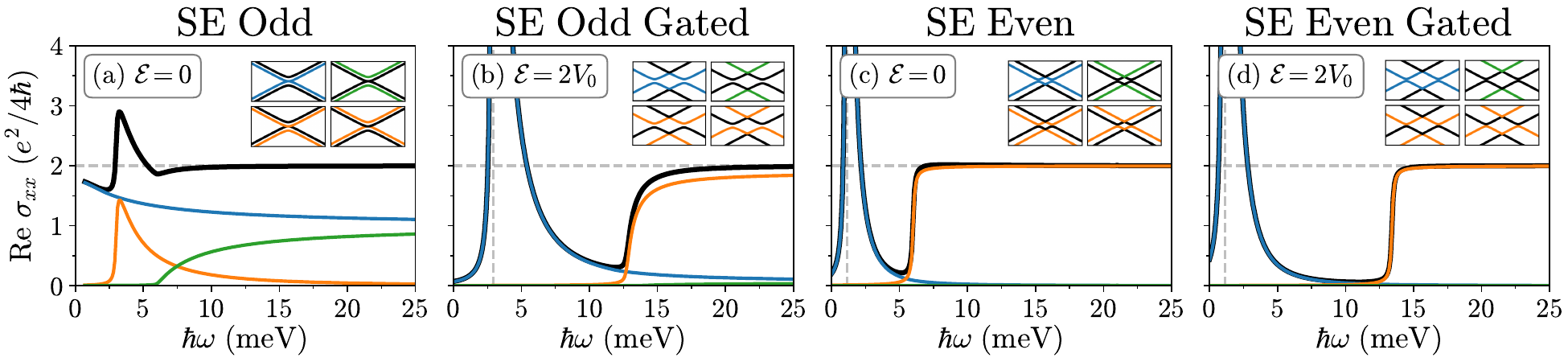}
    \caption{\label{fig:conductivity}
    Optical conductivity can be used to determine the coupling of the offset Dirac cones as a function of twist. The gapped systems exhibit a power-law divergence at the band edge, the gapless system has a finite DC conductivity, and all systems asymptote to twice the optical conductivity of the monolayer in the high frequency limit. Band structures of the systems and interband transitions contributing to the optical conductivity are shown in the figure insets.  Vertical dashed lines correspond to positions where the optical conductivity diverges. \textbf{(a)} the peak onsets at $V_0$ and the location is weakly dependent on scattering and temperature; universal behavior onsets at $2V_0$, \textbf{(b)} the peak is at $V_0\sqrt{\epsilon/(1+\epsilon)}$ where $\epsilon=4\mathcal{E}^2/V_0^2$ and universal behavior onsets near $2\mathcal{E}$, \textbf{(c)} the peak is at $2V_0\sin((\varphi-\theta)/2)$ where $\varphi=\pi/3$ and $\theta=38.21^\circ$ is the twist angle; universal behavior onsets at $2V_0$, \textbf{(d)} The divergent peak is at the same location as in (c), but the onset of universal behavior near $2\mathcal{E}$.}
\end{figure*}
        
        As shown by one of us in Ref. \onlinecite{mele2010commensuration}, the low-energy continuum model for these systems are those of layer Dirac cones coupled by interlayer coupling terms. Explicitly,
        \begingroup
            \setlength\arraycolsep{0.3pt}
        \begin{align}
            H_{\bm{k}}^\text{odd} = \begin{pmatrix}
            \mathcal{E} & k_x - ik_y & V_0 & 0\\
            \!k_x + ik_y & \mathcal{E} & 0 & 0\\
            V_0 & 0 & -\mathcal{E} & \!\!-e^{i\theta}(k_x+ik_y)\!\\
            0&0&\!\!-e^{-i\theta}(k_x-ik_y)\!\!&-\mathcal{E}
            \end{pmatrix}\nonumber\\
            H_{\bm{k}}^\text{even} = \begin{pmatrix}
            \mathcal{E} & k_x - ik_y & V_0 e^{i\varphi/2} & 0\\
            \!k_x + ik_y & \mathcal{E} & 0 & V_0 e^{-i\varphi/2}\\
            V_0 e^{-i\varphi/2} & 0 & -\mathcal{E} & e^{-i\theta}(k_x-ik_y)\!\\
            0&V_0 e^{i\varphi/2}\,\,&e^{i\theta}(k_x+ik_y)\!\!&-\mathcal{E}
            \end{pmatrix}
        \end{align}
        \endgroup 
        whose band structures we superimpose on the band structures obtained from the tight-binding model in Fig. \ref{fig:structures} (e) and (h) respectively. We color these by the band character which exhibits a band inversion; the gap in the SE-even case hosts topological crystalline insulator states \cite{kindermann2015topological}. $\mathcal{E}$ is the electric potential, $V_0$ is the interlayer coherence, $\varphi=\pi/3$ is the pseudospin rotation angle \cite{weckbecker2016low}, and we work in units where $\hbar v_F=1$, and $v_F$ is the Fermi velocity.

    \section{Low Energy Optical Response}
        The low-energy band structures obtained in the last section deviate substantially from the linear band crossing present in monolayer and interlayer decoherent bilayer graphene so this difference should be present in the terahertz optical response. In particular, we focus on the optical conductivity $\sigma$ which we calculate using the Kubo formula \cite{mahan2000many}:
		\begin{align}\label{eq:kubo-sum}
		\sigma_{\mu\nu}(\omega) = \frac{i\hbar}{V} \sum_{\bm{k},s,s'} \frac{f(\varepsilon_s)-f(\varepsilon_{s'})}{\varepsilon_{s'}-\varepsilon_s} \frac{\langle \psi_s|j_\mu|\psi_{s'}\rangle\langle \psi_{s'} |j_\nu|\psi_s\rangle}{\hbar\omega - (\varepsilon_{s'}-\varepsilon_s) + i\eta}
		\end{align}
		where $\omega$ is the frequency of the light, $V$ is the sample volume, $f$ is the Fermi-Dirac distribution function, $T=0$, $|\psi_s\rangle$ and $\varepsilon_s$ are wavefunctions and energies, $j_\mu=(e/\hbar)\partial_{k_\mu} H$ is the current operator, and $\eta$ is a phenomenological scattering rate that we take as $0.1$ meV.
		
		We find the diagonal elements of the conductivity tensor $\sigma_{xx}=\sigma_{yy}$ by numerically evaluating this formula for the SE-even and SE-odd continuum model.  We plot the results in Fig. \ref{fig:conductivity}.  We find that in the SE-odd case, the conductivity remains finite for all frequencies, asymptotes to a finite value at zero frequency, and approaches $e^2/2\hbar$: twice the universal value of the monolayer at large frequencies. This asymptote is accompanied by an equal contribution from the inner and outer two bands, corresponding to the layer hybridization of this system. In an electric field, the SE-odd system gaps and exhibits a power-law divergence at the band edge. In the strong-field limit the Dirac cones of the two layers become independent and the interlayer hybridization disappears; this is accompanied by a selection rule that prevents transitions between different Dirac cones. The gap saturates at $V_0$.
		In SE-even systems, the band structure is always gapped and there is a power-law divergence at the band edge. At low energies, states are strongly hybridized between layers, while at high energies the two Dirac cones are unhybridized and transitions between different Dirac cones only occur at a rate set by the interlayer coherence; this still holds when an electric field is present.
		
		The presence or absence of a divergence in the optical conductivity at finite frequency can be used to distinguish SE-odd and SE-even systems, and its location can be used to determine the interlayer coherence $V_0$. In monolayer graphene, impurities and vacancies lead to low energy peaks, so additional peaks may be present in chemically doped or defect heavy bilayers, but these peak locations cannot be tuned with an applied field \cite{yuan2011optical,houmad2015optical}. For doped systems, Pauli blocking is circumvented and there is an intraband Drude peak near zero frequency.

    \section{Layer Decoupling via Electric Field}
        In the presence of an electric field the two layers are biased by an electrostatic potential $\mathcal{E}$. 
        In the SE-odd case this interlayer bias leads to a gap that is tunable by the electric field, while in the SE-even case, the gap is independent of the electric field.
        In both cases the electric field moves the avoided crossing from the $K$ point to a ring around the $K$ point, and the band structure tends to that of two decoupled Dirac cones separated in energy space by the interlayer potential difference.
        
        In Bernal stacked bilayer graphene, the possibility of opening a gap with electric fields of 1 V/nm were celebrated \cite{castro2007biased}, meanwhile here a gap opens at a voltage that is two orders of magnitude smaller (based on a tight-binding calculation $V_0=338$ meV in the Bernal case). Additionally, the entrance to the strong-field limit is accompanied by an inversion in band character as shown in Fig. \ref{fig:structures} panels (e) evolves to (i) and (h) evolves to (l).
    
    \section{Outlook}
        Twisted bilayer graphene is a rich platform to realize interesting low energy physics, and this physics need not be limited to the magic and small-twist angle regime. At large twist angles and low energies, interlayer coherence effects are essential to describing the electronic structure, and as we have shown the interlayer coherence is directly measurable in terms of the location of a divergence in the optical conductivity.
        
        In SE-odd commensurate twisted bilayer graphene the low energy structure is the same as that of Bernal graphene bilayers except at an energy scale that is two orders of magnitude smaller for the smallest commensurate unit cell. This means that SE-odd commensurate twisted bilayer graphene is a small gap semiconductor that whose gap is tunable through the terahertz region.
        
        The distinctive features of SE-odd and SE-even commensurate twisted bilayer graphene's optical conductivity, including their behavior under electrostatic gating, should be possible to measure on samples with relatively uniform twist angles. Even if the twist angle is not exactly a commensurate angle, we expect domain walls will form between regions of uniform twist angle.
        
    
    \section{Acknowledgments}
        S.T. acknowledges support from the NSF under Grant No. DGE-1845298. E.J.M. acknowledges support from the DOE under Grant No. DE-FG02-84ER45118.

%

	\newpage

	\supplement
	
	\renewcommand{\section}[1]{\begin{center}\large \textbf{#1}\end{center}}
	\widetext
	\setlength{\parindent}{0pt}
	
	\section{Supplementary Information:\\ Electric Field Tunable Band Gap in Commensurate Twisted Bilayer Graphene}
	
	\subsection{Tight-Binding Model}\label{section:tb}
	
	We use the two-center Slater-Koster type model developed and used in \cite{trambly2010localization,moon2013optical}:
	\begin{align}
		t(\vec{r}) &= \begin{cases} V_{pp\pi}^0 e^{-(|\vec{r}|-a_0)/\delta_0} \left(1-\left(\frac{\vec{r}\cdot e_z}{|\vec{r}|}\right)^2\right) + V_{pp\sigma}^0 e^{-(|\vec{r}|-d_0)/\delta_0}\left(\frac{\vec{r}\cdot e_z}{|\vec{r}|}\right)^2, & |\vec{r}|\leq 4a_0\\
			0, & |\vec{r}|>4a_0
		\end{cases}
	\end{align}
	Where the parameters are given by
	\begin{align}
		e_z &= (0,0,1)\\
		a_0 &= a/\sqrt{3} = 0.142 \text{ nm}\\
		d_0 &= 0.335 \text{ nm}\\
		\delta_0 &= 0.184 a = 0.0453 \text{ nm}\\
		V_{pp\pi}(|\vec{r}|) &= 2.7\text{ eV}\\
		V_{pp\sigma}(|\vec{r}|) &= -0.48\text{ eV}
	\end{align}
	For the $21.79^\circ$ twisted bilayer with no interlayer shift and no relaxation, the atomic positions are (in nm)
	\begin{verbatim}
		[0.        , 0.        , 0.        ],        [0.21304225, 0.123     , 0.        ],
		[0.21304225, 0.369     , 0.        ],        [0.4260845 , 0.246     , 0.        ],
		[0.4260845 , 0.492     , 0.        ],        [0.63912675, 0.369     , 0.        ],
		[0.63912675, 0.615     , 0.        ],        [0.07101408, 0.123     , 0.        ],
		[0.28405633, 0.246     , 0.        ],        [0.49709858, 0.123     , 0.        ],
		[0.28405633, 0.492     , 0.        ],        [0.49709858, 0.369     , 0.        ],
		[0.49709858, 0.615     , 0.        ],        [0.71014083, 0.492     , 0.        ],
		[0.        , 0.        , 0.335     ],        [0.15217304, 0.19328571, 0.335     ],
		[0.39564989, 0.15814286, 0.335     ],        [0.30434607, 0.38657143, 0.335     ],
		[0.54782293, 0.35142857, 0.335     ],        [0.45651911, 0.57985714, 0.335     ],
		[0.69999596, 0.54471429, 0.335     ],        [0.26376659, 0.10542857, 0.335     ],
		[0.17246277, 0.33385714, 0.335     ],        [0.41593963, 0.29871429, 0.335     ],
		[0.65941649, 0.26357143, 0.335     ],        [0.32463581, 0.52714286, 0.335     ],
		[0.56811266, 0.492     , 0.335     ],        [0.7202857 , 0.68528571, 0.335     ]
	\end{verbatim}
        
\end{document}